\documentclass[runningheads]{llncs}

\usepackage{graphicx}

\usepackage[hidelinks]{hyperref}

\usepackage{tikz}
\newcommand*\circled[1]{\tikz[baseline=(char.base)]{
            \node[shape=circle,draw,inner sep=1pt] (char) {#1};}}

\begin{document}

\title{Visual Analysis of Ontology Matching Results with the MELT Dashboard}

\titlerunning{Visual Analysis of Ontology Matching Results}

\author{Jan Portisch\inst{1,2}$^\star$ \orcidID{0000-0001-5420-0663} \and
Sven Hertling\inst{1}\thanks{equal contribution} \orcidID{0000-0003-0333-5888} \and
Heiko Paulheim\inst{1}\orcidID{0000-0003-4386-8195}}

\authorrunning{J. Portisch et al.}

\institute{Data and Web Science Group, University of Mannheim, Germany\\
	\email{\{jan, sven, heiko\}@informatik.uni-mannheim.de} \and
	SAP SE Product Engineering Financial Services, Walldorf, Germany\\
\email{\{jan.portisch\}@sap.com}	
}
\maketitle

\begin{abstract}
In this demo, we introduce \textit{MELT Dashboard}, an interactive Web user interface for ontology alignment evaluation which is created with the existing \textit{Matching EvaLuation Toolkit (MELT)}.
Compared to existing, static evaluation interfaces in the ontology matching domain, our dashboard allows for interactive self-service analyses such as a drill down into the matcher performance for data type properties or into the performance of matchers within a certain confidence threshold. In addition, the dashboard offers detailed group evaluation capabilities that allow for the application in broad evaluation campaigns such as the \textit{Ontology Alignment Evaluation Initiative (OAEI)}. 

\end{abstract}

\section{Introduction}
The \textit{Matching EvaLuation Toolkit (MELT)}\footnote{\url{https://github.com/dwslab/melt}} \cite{melt_paper} is an open (MIT-licensed) Java framework for ontology matcher development, tuning, evaluation, and packaging, which integrates well into the existing ontology alignment evaluation infrastructure used by the community, i.e. \textit{SEALS}\footnote{\url{http://www.seals-project.eu}} \cite{garcia-castro_towards_2010,wrigley_semantic_2012} and \textit{HOBBIT}\footnote{\url{http://project-hobbit.eu}} \cite{ngomo_hobbit_2016}. While those frameworks offer programmatic tooling to evaluate ontology matching systems, advanced analyses have to be specifically implemented. Similarly, alignment results are typically presented in the form of static tables which do not allow to explore the actual data. 

\section{Related Work}
The \textit{Alignment API} \cite{alignment_api} is the most well-known ontology matching framework. It allows to develop and evaluate ontology matchers and to render matching results, for example as a \LaTeX{ } figure. The \textit{Semantic Evaluation at Large Scale (SEALS)} framework allows to package matching systems and also provides an evaluation runtime which is capable of calculating precision, recall, and $F_1$. The more recent \textit{Holistic Benchmarking of Big Linked Data (HOBBIT)} runtime works in a similar fashion. In terms of visualization, \textit{Alignment Cubes} \cite{alignmentCubes} allow for a fine grained, interactive visual exploration of alignments. Another framework for working with alignment files is \textit{VOAR} \cite{voar2017} which is a Web-based system where users can upload ontologies and alignments which are then rendered.

Compared to existing work, \textit{MELT Dashboard} is the first interactive Web UI for analyzing and comparing multiple matcher evaluation \emph{results}. The dashboard is particularly helpful for exploring correct and wrong correspondences of matching systems and is, therefore, also suitable for matcher development and debugging.

\section{Architecture}
The dashboard can be used for matchers that were developed in \textit{MELT} but also allows for the evaluation of external matchers that use the well-known alignment format of the \textit{Alignment API}. It is implemented in Java and is included by default in the \textit{MELT} 2.0 release which is available through the maven central repository\footnote{\url{https://mvnrepository.com/artifact/de.uni-mannheim.informatik.dws.melt}}. The \texttt{DashboardBuilder} class is used to generate an HTML page. Without further parameters, a default page can be generated that allows for an in-depth analysis. Alternatively, the dashboard builder allows to completely customize a dashboard before generation -- for instance by adding or deleting selection controls and display panes. After the generation, the self-contained Web page can be viewed locally in the Web browser or be hosted on a server. The page visualization is implemented with \textit{dc.js}\footnote{\url{https://dc-js.github.io/dc.js/}}, a JavaScript charting library with \textit{crossfilter}\footnote{\url{http://crossfilter.github.io/crossfilter/}} support. Once generated, the dashboard can be used also by non-technical users to analyze and compare matcher results.

As matching tasks (and the resulting alignment files) can become very large, the dashboard was developed with a focus on performance. For the \textit{OAEI 2019 KnowledgeGraph} track \cite{dbkwik,kgtrack2020}, for instance, more than 200,000 correspondences are rendered and results are recalculated on the fly when the user performs a drill-down selection. 

\section{Use Case and Demonstration}
One use case for the framework are OAEI campaigns. The \textit{Ontology Alignment Evaluation Initiative} is running evaluation campaigns \cite{oaei_6_years} every year since 2005. Researchers submit generic matching systems for predefined tasks (so called \textit{tracks}) and the track organizers post the results of the systems on each track. The results are typically communicated on the OAEI Web page in a static fashion through one or more tables.\footnote{For an example, see the \textit{Anatomy Track} results page 2019: \url{http://oaei.ontologymatching.org/2019/results/anatomy/index.html}}

In order to demonstrate the capabilities of the dashboard, we generated pages for the following tracks: \textit{Anatomy}, \textit{Conference}, and \textit{KnowledgeGraph}. We included the first two tracks in one dashboard\footnote{Demo link: \url{https://dwslab.github.io/melt/anatomy_conference_dashboard.html}} to show the multi-track capabilities of the toolkit. The \textit{KnowledgeGraph} dashboard\footnote{Demo link: \url{http://oaei.ontologymatching.org/2019/results/knowledgegraph/knowledge_graph_dashboard.html}} was officially used in the OAEI 2019 campaign and shows that the dashboard can handle also combined schema and instance matching tasks at scale. 
The code to generate the dashboards is available in the \texttt{example} folder of the \textit{MELT} project.\footnote{\url{https://github.com/dwslab/melt/tree/master/examples/meltDashboard}} It can be seen that few lines of code are necessary to generate comprehensive evaluation pages.

\begin{figure}
\centering
\includegraphics[scale=0.5]{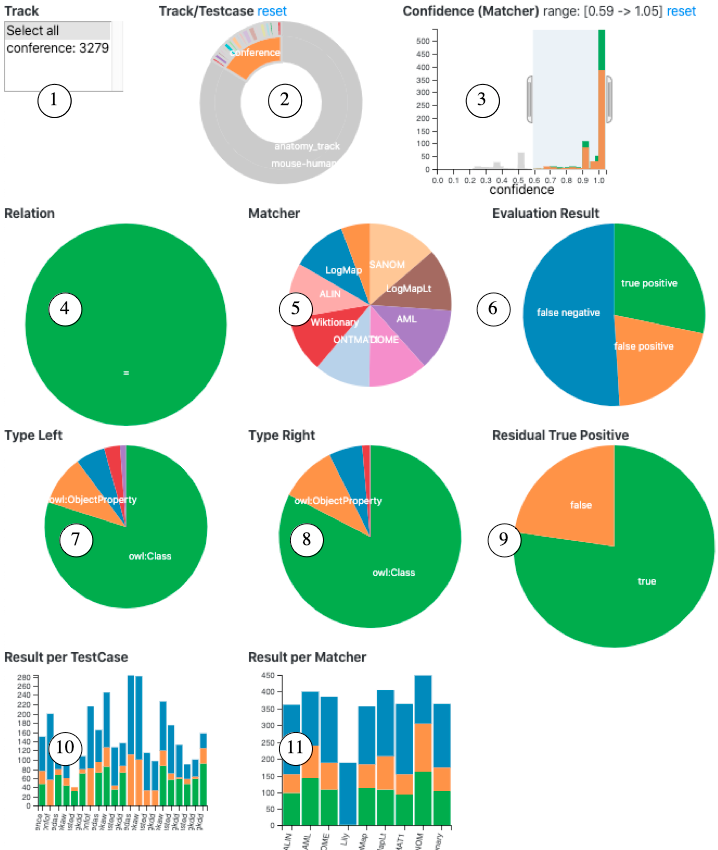}
\caption{Dashboard for the \textit{OAEI Anatomy/Conference Tracks}. The numbered controls are clickable to drill down into the data. If clicked, all elements change automatically to reflect the current selection.}
\label{fig:dashboard_numbered}
\end{figure}

An annotated screenshot of the controls for the \textit{Anatomy/Conference} dashboard is depicted in Figure \ref{fig:dashboard_numbered}. Each numbered element is clickable in order to allow for a sub-selection. For example, in element \circled{2}, the \textit{Conference} track has been selected and all elements in the dashboard show the results for this subselection. The controls in the given sample dashboard are as follows: \circled{1} selection of the track, \circled{2} selection of the track/test case (the \textit{Conference} track is selected with all test cases), \circled{3} confidence interval of the matchers (an interval of $[0.59, 1.05]$ is selected), \circled{4} relation (only equivalence for this track), \circled{5} matching systems, \circled{6} the share of true/false positives (TP/FP) and false negatives (FN), \circled{7}/\circled{8} the type of the left/right element in each correspondence (e.g., class, object property, datatype property), \circled{9} the share of residual true positives (i.e., non-trivial correspondences generated by a configurable baseline matcher), \circled{10} the total number of correspondences found per test case -- the performance result of each match (TP/FP/FN) is color coded, and \circled{11} the color-coded correspondences found per matcher.\\
Below the controls, the default dashboard shows the performance results per matcher, i.e. micro and macro averages of precision (P), recall (R), and F-score ($F_1$) in a table as well as concrete correspondences in a further table (both are not shown in Figure \ref{fig:dashboard_numbered}). The data and all controls are updated automatically when a selection is performed. For example, if the \textit{Anatomy} track is selected (control \circled{2}) for matcher \textit{Wiktionary} \cite{wiktionary_matcher} (control \circled{5}), and only false negative correspondences (control \circled{6}) are desired, the correspondence table will show examples of false negative matches for the \textit{Wiktionary} matching system on the \textit{Anatomy} track.

\section{Conclusion and Future Work}
In this paper, we presented the \textit{MELT Dashboard}, an interactive Web user interface for ontology alignment evaluation.
The tool allows to generate dashboards easily and to use them for a detailed evaluation in a drill-down fashion. With the new functionality, we hope to increase the transparency and the understanding of matching systems in the ontology alignment community and to make in-depth evaluation capabilities available to a broader audience without the need of installing any software. The first usage in the OAEI 2019 campaign showed that the dashboard can be used for broad evaluation campaigns of multiple matchers on multiple matching tasks. In the future, we plan to extend the interface with further controls, to make it more visually appealing, and to grow its adoption.

\bibliographystyle{splncs04}
\bibliography{references}

\begin{thebibliography}{10}
\providecommand{\url}[1]{\texttt{#1}}
\providecommand{\urlprefix}{URL }
\providecommand{\doi}[1]{https://doi.org/#1}

\bibitem{alignment_api}
David, J., Euzenat, J., Scharffe, F., dos Santos, C.T.: The alignment {API}
  4.0. Semantic Web  \textbf{2}(1),  3--10 (2011)

\bibitem{oaei_6_years}
Euzenat, J., Meilicke, C., Stuckenschmidt, H., Shvaiko, P., dos Santos, C.T.:
  Ontology alignment evaluation initiative: Six years of experience. J. Data
  Semantics  \textbf{15},  158--192 (2011)

\bibitem{garcia-castro_towards_2010}
García-Castro, R., Esteban-Gutiérrez, M., Gómez-Pérez, A.: Towards an
  infrastructure for the evaluation of semantic technologies. In: {eChallenges}
  e-2010 {Conference}. pp.~1--7 (2010)

\bibitem{dbkwik}
Hertling, S., Paulheim, H.: {DBkWik}: A consolidated knowledge graph from
  thousands of wikis. In: {IEEE} International Conference on Big Knowledge,
  {ICBK}. pp. 17--24 (2018)

\bibitem{kgtrack2020}
Hertling, S., Paulheim, H.: The knowledge graph track at {OAEI} - gold
  standards, baselines, and the golden hammer bias. In: The Semantic Web - 17th
  International Conference, ESWC (2020), [to appear]

\bibitem{melt_paper}
Hertling, S., Portisch, J., Paulheim, H.: {MELT} - matching evaluation toolkit.
  In: Semantic Systems. The Power of {AI} and Knowledge Graphs - 15th
  International Conference, SEMANTiCS. pp. 231--245 (2019)

\bibitem{alignmentCubes}
Ivanova, V., Bach, B., Pietriga, E., Lambrix, P.: Alignment cubes: Towards
  interactive visual exploration and evaluation of multiple ontology
  alignments. In: International Semantic Web Conference ({ISWC}). pp. 400--417
  (2017)

\bibitem{ngomo_hobbit_2016}
Ngomo, A.C.N., Röder, M.: {HOBBIT}: Holistic benchmarking for big linked data.
  ERCIM News  \textbf{105} (2016)

\bibitem{wiktionary_matcher}
Portisch, J., Hladik, M., Paulheim, H.: {Wiktionary} matcher. In: 14th
  International Workshop on Ontology Matching co-located with the 18th
  International Semantic Web Conference ({ISWC}). pp. 181--188 (2019)

\bibitem{voar2017}
Severo, B., Trojahn, C., Vieira, R.: {VOAR} 3.0 : a configurable environment
  for manipulating multiple ontology alignments. In: International Semantic Web
  Conference (Posters, Demos {\&} Industry Tracks). {CEUR} Workshop
  Proceedings, vol.~1963 (2017)

\bibitem{wrigley_semantic_2012}
Wrigley, S.N., García-Castro, R., Nixon, L.: Semantic evaluation at large
  scale ({SEALS}). In: 21st international conference companion on {World}
  {Wide} {Web} - {WWW}. pp. 299--302 (2012)

\end{thebibliography}

\end{document}